\documentclass[11pt]{article}
\usepackage{jheppub}

\usepackage{amsmath,amssymb,mathtools,bm}
\usepackage{microtype}
\usepackage{comment}
\usepackage{enumitem}

\hypersetup{colorlinks=true,linkcolor=blue,citecolor=blue,urlcolor=blue,
            bookmarksnumbered=true,breaklinks=true}
\usepackage{orcidlink}
\newcommand{\GB}{\mathcal G}
\newcommand{\dd}{\mathrm d}
\newcommand{\Arg}{\operatorname{Arg}}
\newcommand{\wh}{\mathrm{wh}}
\newcommand{\IDB}{\mathrm{IDB}}
\newcommand{\eps}{\epsilon}

\title{ Scalar Gauss-Bonnet Wormholes and the Imaginary Distance Bound   }

\author[a]{Alex Kehagias
\orcidlink{0000-0001-6080-6215},
}
\author[b]{Antonio~Riotto
\orcidlink{0000-0001-6948-0856}
}
\affiliation[a]{Physics Division, National Technical University of Athens, Athens, 15780, Greece}
\affiliation[b]{Department of Theoretical Physics and Gravitational Wave Science Center,  \\
24 quai E. Ansermet, CH-1211 Geneva 4, Switzerland}

\abstract{
We study $O(4)$ symmetric Euclidean wormholes in four-dimensional gravity with a massless scalar linearly coupled to the Gauss-Bonnet invariant. At nonzero coupling, the Einstein wormhole becomes a regular complex saddle that satisfies the Kontsevich--Segal--Witten criterion until its branch ends at a finite coupling. In asymptotically flat space, the Gauss-Bonnet term modifies the scalar displacement and the on-shell action for finite wormholes, but the size dependent parts of both corrections vanish in the large throat limit. The original imaginary distance threshold therefore remains unchanged. A negative cosmological constant already lowers the Einstein large throat imaginary distance relative to its flat space value. When the Gauss-Bonnet coupling is also present, the scalar is no longer a modulus as it develops a logarithmic running and its trajectory has infinite length. Subtracting the logarithmic asymptotic behavior leaves a finite regulated displacement. Its large throat value is shifted upward from the Einstein AdS result by the Gauss-Bonnet interaction and determines the leading large charge behavior of the renormalized fixed charge action.
}

\emailAdd{kehagias@central.ntua.gr}
\emailAdd{Antonio.Riotto@unige.ch}
\begin{document}
\maketitle
\flushbottom

%%%%%%%%%%%%%%%%%%%%%%%%%%%%%%%%%%%%%%%%%%%%%%%%%%%%%%%%%%%%%%%%%%%%%

\section{Introduction}

Euclidean wormholes are  solutions of the Euclidean gravitational field equations
possessing more than one asymptotic region.  They  play a significant role for the understanding of   quantum
gravity ~\cite{Giddings:1987cg,Coleman:1988cy}, in particular  in connection with the black hole information problem,
the late time behaviour of the spectral form factor and the statistics of
black hole microstates~\cite{Saad:2018bqo,Saad:2019lba,Penington:2019kki,%
Almheiri:2019qdq}. 

Euclidean wormholes generate connected contributions between otherwise
disconnected boundaries, and therefore, the corresponding partition function need not be 
factorized. Such contributions are often interpreted as arising from an
ensemble average over effective coupling constants
\cite{Coleman:1988cy,Saad:2019lba,Saad:2021rcu}. This interpretation however, is
problematic for a fixed string theory vacuum, where the effective couplings
are determined by  compactification rather than averaging. The status of the corresponding saddle points in AdS/CFT has
been repeatedly scrutinized~\cite{Maldacena:2004rf,Marolf:2021kjc,%
Maloney:2025tnn}, and it remains an open question which of them should be
retained in the path integral.

Simple wormholes are supported by a massless scalar
field. The Einstein equations sourced by a real scalar do not admit a throat,
and the standard construction proceeds either by dualizing the scalar into a
three form field strength, or, equivalently at the level of the equations of
motion, by continuing the scalar to purely imaginary values, which flips the
sign of its contribution to the stress tensor~\cite{Giddings:1987cg,%
Arkani-Hamed:2007cpn}. Recently, these wormholes have been used to argue for an upper bound
on how far scalar boundary values can be analytically continued in
the imaginary direction~\cite{DiUbaldo:2026rly,Maldacena:2026jqd}. In the Dirichlet problem the asymptotic values $\varphi_\infty$
of the scalars are the coupling constants of the gravitational theory, and,
in AdS, of the dual conformal field theory. The existence of an on-shell
wormhole at a given imaginary separation of the two endpoint values is then
read as the statement that the analytic continuation of the theory in these
couplings has broken down. The resulting imaginary distance bound  is an
upper bound on how far $\mathrm{Im}\,\varphi_\infty$ can be continued in any
consistent theory of quantum gravity.  In four dimensional flat space, and in
Planck units in which the scalar kinetic term is normalized,  the threshold is $\pi\sqrt3/2$. The bound is conjectured
to be enforced, in any given ultraviolet completion, by effects of the form
$c_ne^{in\varphi}$ which are bounded for real $\varphi$ but become
uncontrolled once $\varphi$ acquires an imaginary part. Upon Kaluza-Klein
reduction the imaginary distance bound reduces to the Kontsevich--Segal--Witten (KSW) allowability
criterion for complex metrics~\cite{Kontsevich:2021dmb,Witten:2021nzp,%
BenettiGenolini:2026raa} and, in the presence of gauge fields, to the weak gravity
conjecture~\cite{Arkani-Hamed:2006emk,Montero:2015ofa,Harlow:2022ich}.
A related argument for axions, based on positivity of the
gravitational path integral, was developed in
Ref.~\cite{DiUbaldo:2026rly}.

The imaginary distance bound is
therefore a candidate unifying principle behind a set of otherwise
independent swampland constraints~\cite{Ooguri:2006in,Grimm:2018ohb}, and it
is natural to ask how robust it is. We should stress that whether these
complex saddles contribute to the path integral at all is not settled, and
that arguments to the contrary have been given
in~\cite{Witten:2026twr,Held:2026bbo,Held:2026bbo,Mallik:2026man}, while their perturbative
stability has been analyzed
in~\cite{Hertog:2018kbz,Loges:2022nuw,Hertog:2024nys}. In what follows we
adopt the point of view of Ref.~\cite{Maldacena:2026jqd} and assume that
they do.

The imaginary distance bound, as formulated in Ref.~\cite{Maldacena:2026jqd},
is a statement about two derivative Einstein--scalar gravity. On the other
hand, gravity coupled to matter is an effective field theory, and any
ultraviolet completion inevitably supplements the Einstein--Hilbert term with
higher curvature operators.
The wormhole throat is precisely the region where the curvature
is largest, the threshold is extracted from a family of solutions labelled by
the shift charge, and the argument leading to the bound is a statement about
the geometry of that family in field space. It is therefore legitimate to ask
whether the threshold is an artefact of the two derivative truncation or a
property that survives the leading corrections to it. 
Higher derivative Gauss-Bonnet corrections to the imaginary distance argument have
 been discussed for supersymmetric axion-saxion wormholes
in Ref.~\cite{Licciardello:2026roi}, where their contribution is
evaluated on solutions of the two derivative theory.
Here, we  focus on the linear coupling of a massless scalar to the
four dimensional Gauss-Bonnet invariant $\GB$, i.e., 
\begin{equation}
 {\cal L}\supset m\,\phi\,\GB, \qquad
\GB=R_{\mu\nu\rho\sigma}R^{\mu\nu\rho\sigma}-4R_{\mu\nu}R^{\mu\nu}+R^2.
\label{eq:GB-I}
\end{equation}
 This operator is singled out for several
reasons. The Gauss-Bonnet density is topological in four dimensions, so it
contributes to the dynamics only through its coupling to the scalar. In addition, the
field equations remain of second order, so that no additional ghost like
degrees of freedom are propagated~\cite{Lovelock:1971yv,Boulware:1985wk}. Let us note that such term
is generated at leading order in the $\alpha'$ expansion of the heterotic
string~\cite{Zwiebach:1985uq,Gross:1986mw,Metsaev:1987zx}. Furthermore, the theory breaks the
shift symmetry of $\phi$ only by a total derivative, so that the shift charge
$\mathcal Q$ which labels the wormhole family is still conserved and the
solutions remain a one parameter family. The linear coupling also supports black holes with scalar
hair~\cite{Sotiriou:2013qea}, providing another setting in which
its effects on curved geometries can be studied.
Euclidean wormholes in related theories with a scalar coupled
to the Gauss-Bonnet invariant have also been studied in
Ref.~\cite{Chew:2020lkj}.
Our purpose here is to determine the fate of the imaginary
distance bound in the presence of a coupling of the form of Eq. \eqref{eq:GB-I}, and our findings may be
summarized as follows.

The first consequence of the Gauss-Bonnet interaction is qualitative. The source
$m\GB$ in the scalar field equation is real on a real metric, so the
configuration in which the metric is real and the scalar is purely imaginary
is no longer a solution. The Einstein wormhole continues instead to a
genuinely complex saddle, whose two exterior regions are exchanged by complex
conjugation and which is joined through a regular throat of real areal
radius. We construct this saddle in a coordinate which is regular across the
throat, and we show that the continuation exists only over a finite range
 of the dimensionless coupling, beyond
which the relevant root of the algebraic constraint merges with another one
and the branch terminates at an algebraic branch point. Equivalently, at
fixed $m$ the throat radius is bounded from below. We prove analytically that
the whole physical branch satisfies the KSW criterion, so that the
obstruction to extend the solution is dynamical and does not originate
from the validity of the complex metric.

Second, for a wormhole with finite throat radius, the leading
Gauss-Bonnet correction increases the imaginary displacement.
The reference subtracted action contains a size dependent
contribution and a topological term determined by the scalar
boundary data, so its sign is not universal. At fixed coupling,
the size dependent corrections vanish in the large throat
limit, leaving the asymptotically flat imaginary distance
threshold unchanged.

Finally, the situation in the presence of a negative cosmological constant is
different. There the Gauss-Bonnet coupling generates a logarithmic running
of the scalar near each conformal boundary, so that the Hermitian length of
the scalar trajectory is infinite and the bound cannot be transcribed
directly in terms of it. The logarithmic term is however common to the two
ends and cancels in their difference, which leaves a finite regulated
endpoint displacement whose large throat limit is shifted with respect to the
Einstein value and controlled by the large charge behaviour of the
renormalized fixed charge action.
Since the Gauss-Bonnet coupling lifts
the scalar modulus in EAdS, the dual coupling runs and the boundary theory is
not conformal. The displacement should therefore be regarded as the imaginary
part of a running coupling rather than as a distance on a conformal
manifold.
The endpoint difference is independent of the logarithmic
subtraction scale, while its interpretation through the fixed
charge action uses the boundary prescription specified below.

The paper is organized as follows. In Section~\ref{sec:setup} we fix the
action, the conventions and the field equations, and we derive the algebraic
constraint that determines the $O(4)$ symmetric wormhole. In
Section~\ref{sec:m0} we review the Einstein limit, which serves both as a
benchmark and as the origin of the coordinates used throughout. Section
\ref{sec:existence} is devoted to the construction of the complex wormhole,
to its behaviour near the throat and to the algebraic branch point which
terminates the branch. In Section~\ref{sec:ksw} we verify the KSW criterion
along the whole physical branch. Section~\ref{sec:scalar} contains the
computation of the endpoint displacement and of the on-shell action, and the
resulting statement about the flat space imaginary distance bound. In
Section~\ref{sec:adsdistance} we repeat the analysis for a negative
cosmological constant and discuss the regulated AdS threshold. We conclude
in Section~\ref{sec:conclusion}. Appendix~\ref{app:action} collects the
boundary terms and the details of the on-shell action.

Throughout the paper we work in Euclidean signature and in four spacetime
dimensions, so that $AdS$ here refers to Euclidean AdS (EAdS). In addition, we set the scalar kinetic normalization as in \eqref{eq:action},
and we denote by $\mathcal Q$ the, generally complex, shift charge.

\section{Action and conventions}
\label{sec:setup}

Let us consider a massless scalar field $\phi$ non-minimally coupled to gravity  with Euclidean action
\begin{equation}
 I_{\rm bulk}[g,\phi]
 =-\frac{1}{16\pi G_N}\int \dd^4x\,\sqrt g\left[
 R-\frac12\nabla_\mu\phi\nabla^\mu\phi+m\phi\GB
 \right],
 \label{eq:action}
\end{equation}
where $\GB$ 
is the Gauss-Bonnet scalar given in Eq. \eqref{eq:GB-I}, and $m$ is a real dimensionfull coupling constant. The boundary terms required by the Dirichlet variational problem will
be included when evaluating the on-shell action in
Section~\ref{sec:scalar} and Appendix~\ref{app:action}. They do not
modify the bulk field equations derived below.
 We have  chosen in the action \eqref{eq:action} a linear coupling because, among possible  interactions of the form
$S(\phi)\GB$, it preserves the scalar shift symmetry at the level of
the bulk field equations.  Indeed, under $\phi\to\phi+c$ the action, including the boundary term
\eqref{eq:GBBoundaryTerm}, changes only by the topological quantity
$-(2\pi mc/G_N)\chi$, where $\chi$ is the Euler characteristic of the
manifold. The field equations are therefore invariant, and the conserved
modified scalar flux introduced in \eqref{eq:scalarfirstintegral} organizes
the wormhole solutions in fixed charge sectors, allowing direct comparison
with the imaginary distance argument of Ref.~\cite{Maldacena:2026jqd}. The
shift does however affect the comparison between on-shell actions of
geometries with different topology, as discussed in Section~\ref{sec:scalar}. The linear coupling also has a natural supersymmetric motivation, since   supersymmetry relates the parity-even $\phi\GB$ to the gravitational Pontryagin parity-odd coupling $aR\widetilde R$, where $a$ is the axion, with both terms arising from the same supersymmetric curvature-squared interaction~\cite{Cecotti:1987mr}.
This supersymmetric structure motivates the interaction in
\eqref{eq:action}, which we study independently of its full
supersymmetric completion.

The scalar field equation  which follow form Eq. \eqref{eq:action} is 
\begin{align}
    \Box\phi+m\GB=0,
 \label{eq:sEOM}
\end{align}
whereas the gravitational field equation can be  written as
    \begin{equation}
G_{\mu\nu}
-\frac{1}{2}\nabla_{\mu}\phi\nabla_{\nu}\phi
+\frac{1}{4}g_{\mu\nu}(\nabla\phi)^{2}
+m\mathcal H_{\mu\nu}=0,
\label{eq:metricEOM}
\end{equation}
with
\begin{align}
\mathcal H_{\mu\nu}
={}&
2R\left(
g_{\mu\nu}\Box \phi-\nabla_{\mu}\nabla_{\nu}\phi
\right)
-4R_{\mu\nu}\Box\phi
-4g_{\mu\nu}R^{\rho\sigma}
\nabla_{\rho}\nabla_{\sigma}\phi
\nonumber\\
&
+4R_{\mu}{}^{\rho}
\nabla_{\rho}\nabla_{\nu}\phi
+4R_{\nu}{}^{\rho}
\nabla_{\rho}\nabla_{\mu}\phi
+4R_{\mu\rho\nu\sigma}
\nabla^{\rho}\nabla^{\sigma}\phi,
\label{eq:Hmn}
\end{align}
 the  Gauss-Bonnet contribution and $G_{\mu\nu}$ the Einstein tensor. 

We are looking for  $O(4)$ symmetric wormhole solutions of the form
\begin{equation}
 \dd s^2=\frac{\dd r^2}{f(r)^2}+r^2\dd\Omega_3^2,
 \qquad \phi=\phi(r),
 \qquad F(r)\equiv f(r)^2,
 \label{eq:ansatz}
\end{equation}
with $r$ the areal radius and $\dd\Omega_3^2$ the round metric on the unit three sphere. 
Then, the scalar equation
is written as
\begin{equation}
\frac{\dd}{\dd r}
\left(r^{3}f\phi'\right)
=-24 m(f^{2}-1)f', 
\end{equation}
which  admits the first integral
\begin{equation}
 r^3f\phi'
 +8m\left(f^3-3f\right)
 =
 \mathcal Q,
 \label{eq:scalarfirstintegral}
\end{equation}
where $\mathcal Q$ is a generally complex constant of integration. Physically $\mathcal Q$ is, up to normalization, the charge associated with the shift symmetry of $\phi$ that is present at the level of the field equation \eqref{eq:sEOM}, deformed by the Gauss-Bonnet coupling.  Solving for $\phi'$ we find
\begin{equation}
 \phi'(r)
 =
 \frac{\mathcal Q}{r^3f(r)}
 -\frac{8m}{r^3}
 \left[f(r)^2-3\right].
 \label{eq:phiprime}
\end{equation}
and, consequently, 
the scalar $\phi$ is given by the integral
\begin{equation}
\phi(r)
=\phi(r_0)+
\int_{r_0}^{r}
\left[-
\frac{8m\bigl(f(k)^2-3\bigr)}{k^{3}}
+\frac{\mathcal{Q}}{k^{3}f(k)}
\right] \dd k. 
\label{eq:phiint}
\end{equation}

From the Einstein equations \eqref{eq:metricEOM} we find that the $rr-$component is written as  \begin{equation}
 \frac{r^3f^2\phi'^2}{2}-6rf^2+6r+24mf^4\phi'-24mf^2\phi'=0.
\label{eq:rr}
\end{equation}
whereas the rest of the equations are satisfied due to the scalar field equation and the Bianchi identity. 
By  using the first integral \eqref{eq:phiprime}, we find that the only non-trivial Einstein equation is written as 
\begin{equation}
\begin{aligned}
0={}&12r^4+\mathcal Q^2
-12(r^4+48m^2)f^2+32m\mathcal Q f^3
\\
&+1152m^2f^4-320m^2f^6.
\end{aligned}
\label{eq:alg}
\end{equation}
This is an algebraic equation for $f$, which once solved, it determined not only the metric but also the scalar through the integral \eqref{eq:phiint}. 

\section{The Einstein limit}
\label{sec:m0}

It is useful to discuss the $m=0$ solution in detail~\cite{Giddings:1987cg}, both as a benchmark for the full $m\neq0$ case and also because the coordinates introduced here will be used throughout the rest of the paper. Setting $m=0$ in \eqref{eq:alg} gives
\begin{equation}
    12r^4+\mathcal{Q}^2-12r^4f_0^2=0
\end{equation}
which is solved by  
\begin{equation}
 F_0(r)
 =1-\left(\frac{r_0}{r}\right)^4,
 \qquad
 r_0^4=\frac{Q^2}{12}, \qquad \mathcal Q=i Q,
 \label{eq:m0metric}
\end{equation}
where $F_0(r)=f_0(r)^2$.
This is precisely the four dimensional case of the wormhole metric~\cite{Giddings:1987cg,Maldacena:2026jqd}. Now we may introduce a coordinate $u\in(-1,1)$ through the relation
\begin{equation}
 r^4=\frac{r_0^4}{1-u^2},
 \label{eq:ucoordinate}
\end{equation}
where $r_0$ is the areal radius of the throat, so that the two asymptotically flat regions sit at $u=\pm1$ and the throat sits at $u=0$. Fixing the sign of $f_0$ consistently across the two exterior regions gives the compact expression
\begin{equation}
 f_0(u)=u,
 \qquad
 F_0(u)=u^2.
\end{equation}
Evaluating then \eqref{eq:phiint} at $m=0$ and normalizing the scalar by $\phi(0)=0$ gives the pure imaginary solution
\begin{equation}
 \phi_0(u)
 =i\frac{Q}{2r_0^2}\arcsin u
 =i\,\operatorname{sgn}(Q)\sqrt3\,\arcsin u.
 \label{eq:m0scalar}
\end{equation}
 The asymptotic values of the scalar are therefore
\begin{equation}
 \phi_{0,\pm}
 =\pm i\,\operatorname{sgn}(Q)\frac{\pi\sqrt3}{2},
 \qquad
 \left|-i\Delta\phi_{0,\infty}\right|=\pi\sqrt3,
 \label{eq:m0endpoints}
\end{equation}
and the half wormhole displacement, defined as the size of the imaginary part of either asymptotic value, is
\begin{equation}
 d_{\wh}(Q,0)=\frac{\pi\sqrt3}{2}.
 \label{eq:m0IDB}
\end{equation}
This is exactly the four dimensional imaginary distance bound found in the original construction, $\tau_{\IDB}=(\pi/2)\sqrt{2(D-1)/(D-2)}$ evaluated at $D=4$. It is against this number that we measure the effect of the Gauss-Bonnet coupling below.

\section{Existence and structure of the complex wormhole}
\label{sec:existence}

We now turn to the solutions of \eqref{eq:phiint} and \eqref{eq:alg} that is continuously connected to the Einstein wormhole of Section~\ref{sec:m0} as $m\to0$. 
Clearly, due to the real source $m \GB$ in the scalar field equation \eqref{eq:sEOM}, there is no pure imaginary solution for the scalar. In this case, the scalar field is generally complex with both real and imaginary parts.

To proceed, let us 
 first locate the candidate throat and study its local structure. We will introduce a coordinate that is regular across the throat and prove that the resulting solution extends smoothly across the whole wormhole for a finite range of the coupling.

\subsection{Location of the throat and its local structure}

As in the Einstein case, we set $\mathcal Q=iQ$ with $Q$ real. This is
the choice compatible with the conjugation symmetry of the solution
discussed below, and it is assumed in the rest of the paper.
The location of the wormhole throat is at $r=r_0>0$ which is defined by $F(r_0)=f(r_0)^2=0$. Setting $f=0$ in \eqref{eq:alg} gives
\begin{equation}
 r_0^4=\frac{Q^2}{12}.
\label{eq:throatradius}
\end{equation}
Therefore, the Gauss-Bonnet coupling does not shift the areal radius of the candidate throat away from its Einstein value. This is a direct consequence of the fact that every term in \eqref{eq:alg} proportional to $m$ also carries a positive power of $f$, so all such terms vanish automatically at $f=0$. A nonzero charge $Q$ remains essential for the throat to sit at positive radius, since $Q=0$ collapses \eqref{eq:throatradius} to $r_0=0$.

The vanishing of $F$ at $r_0$ is necessary but not sufficient for a genuine throat, since it could equally well signal a curvature singularity if the zero is not simple in an appropriate sense. To settle this we expand \eqref{eq:alg} locally,  define
\begin{equation}
 t=r^4-r_0^4,
 \qquad
 D=r_0^4+48m^2>0,
 \label{eq:tDdefinitions}
\end{equation}
and then look for a solution of the form $f=c_1t^{1/2}+c_2t+O(t^{3/2})$ near $t=0$, consistent with $F=f^2$, which vanish at $t=0$. Substituting into \eqref{eq:alg} and matching powers of $t^{1/2}$ order by order, fixes both coefficients uniquely on the branch with
$\operatorname{Re}f>0$ for $r$ slightly larger than $r_0$,
\begin{equation}
 f(t)
 =\frac{t^{1/2}}{D^{1/2}}
 +\frac{4imQ}{3D^2}t
 +O(t^{3/2}),
 \label{eq:localf}
\end{equation}
and correspondingly
\begin{equation}
 F(t)
 =\frac{t}{D}
 +\frac{8imQ}{3D^{5/2}}t^{3/2}
 +O(t^2).
 \label{eq:localF}
\end{equation}
The leading behavior $F\propto t$ is exactly what a smooth minimum of the sphere radius requires, with the two square root branches of $t^{1/2}$ in \eqref{eq:localf} corresponding to the two sides of the wormhole. Differentiating \eqref{eq:localF} at $t=0$ gives the derivative of $F$ at the throat,
\begin{equation}
 F'(r_0)=\frac{4r_0^3}{r_0^4+48m^2}>0,
 \label{eq:FprimeThroat}
\end{equation}
which is manifestly positive for all real $m$.  Thus,  the sphere radius increases monotonically on both sides of $r=r_0$ near the throat, exactly as it should for a genuine wormhole throat rather than a more singular degeneration of the geometry.

Note that the curvature invariants remain finite at $r_0$, and the divergence of $g_{rr}=1/F$ at $r=r_0$, where $F(r_0)=0$, is merely a coordinate artifact of the areal radius chart rather than a genuine physical singularity. Indeed, the Ricci scalar and the Gauss–Bonnet invariant are given by\begin{equation}R(r_0)=\frac{6}{r_0^2}-\frac{12r_0^2}{r_0^4+48m^2},\qquad\GB(r_0)=-\frac{48}{r_0^4+48m^2},\label{eq:throatcurvatures}\end{equation}both of which are finite for any real $m$ and $r_0>0$. This apparent singularity at $r_0$ is removed below by passing to the regular global $u$-coordinate.

\subsection{A global coordinate and the complex wormhole}

To describe both exterior regions in a single coordinate system we use the coordinate $u$  to express the areal radius as
\begin{align}
 r(u)&=r_0(1-u^2)^{-1/4},
 \qquad -1<u<1.
 \label{eq:rOfu}
 \end{align}
 In addition, we parametrize $f$ as 
 \begin{align}
 f(u)&=u\,h(u),
 \label{eq:hdefinition}
 \end{align}
 and we introduce the dimensionless parameter
 \begin{align}
 \eps&=\frac{2m}{Q},
\end{align}
which  measures the strength of the Gauss
Bonnet coupling relative to the wormhole charge. Substitution into
\eqref{eq:alg} gives
\begin{equation}
\begin{aligned}
G(u,h,\eps)
={}&1-h^2
+16i\eps u(1-u^2)h^3
\\
&+\eps^2(1-u^2)
\left[
-144h^2+288u^2h^4-80u^4h^6
\right]
=0.
\end{aligned}
\label{eq:G}
\end{equation}
At zero coupling Eq.~\eqref{eq:G} reduces to
\begin{equation}
 G(u,h,0)=1-h^2, 
\end{equation}
and the physical root connected to the Einstein solution is therefore
\begin{equation}
 h(u)=1,
 \qquad
 \left.
 \frac{\partial G}{\partial h}
 \right|_{\epsilon=0,\,h=1}
 =-2.
 \label{eq:G0}
\end{equation}
The nonzero derivative selects a unique branch connected continuously
to the Einstein wormhole for sufficiently small $|\eps|$. This branch
can be followed across the full interval until it reaches the
branch point discussed below. Notice that since
\begin{equation}
G(u,0,\eps)=1,
\end{equation}
$h$ cannot vanish anywhere along the branch, and  consequently,
\begin{equation}
F(u)=u^2h(u)^2
\end{equation}
has the required double zero at $u=0$ and no additional zero in either
exterior region.
For real $m$ and $Q$, Eq.~\eqref{eq:G} obeys
\begin{equation}
G(-u,h^*,\eps)=G(u,h,\eps)^*,
\end{equation}
so that continuity from $h(u)=1$ therefore gives
\begin{equation}
h(-u)=h(u)^*.
\label{eq:hsymmetry}
\end{equation}
The two exterior geometries are thus complex conjugates of one another.

At the throat and at the two asymptotic ends one finds
\begin{equation}
h(0)=\frac{1}{\sqrt{1+144\eps^2}},
\qquad
h(\pm1)=1.
\label{eq:hendpoints}
\end{equation}
The positive endpoint root is selected by continuity from the Einstein
solution. The throat value is real and nonzero for every real $\eps$,
while $F\to1$ at both ends. The geometry therefore has a regular
minimal sphere at $u=0$ and remains asymptotically flat throughout the
domain of the physical branch. Although the metric is generally
complex in the interior, it becomes real both at the throat and at the
two asymptotic ends.

The regularity of the throat can also be seen directly from the metric
in the coordinate $u$. 
In terms of $u$, 
the two-sided metric becomes
\begin{equation}
\dd s^2
=
\frac{r(u)^2}
{4(1-u^2)^2h(u)^2}\,\dd u^2
+r(u)^2\dd\Omega_3^2 ,
\label{eq:metricU}
\end{equation}
so that at the throat, its radial component is
\begin{equation}
g_{uu}(0)
=
\frac{r_0^2}{4h(0)^2}
=
\frac{r_0^2}{4}
\left(1+144\eps^2\right)>0.
\label{eq:guuThroat}
\end{equation}
In addition, we find that
\begin{equation}
r'(0)=0,
\qquad
r''(0)=\frac{r_0}{2}>0,
\end{equation}
which confirms that the areal radius has a smooth minimum at $u=0$.
Since $h(u)$ remains analytic and nonzero throughout the domain of the
physical branch, the metric is analytic and nondegenerate at the throat.

\subsection{The limiting branch point}

As $|\eps|$ is increased from zero, the physical branch obtained by
continuous continuation of the Einstein root $h=1$ may cease to exist when
it merges with another root of \eqref{eq:G}. Such a coincidence  produces a
double root and is characterized by
\begin{equation}
G(u,h,\eps)=0,
\qquad
\partial_hG(u,h,\eps)=0.
\label{eq:doubleRoot}
\end{equation}
The coincidence of the roots cannot occur at the throat since setting $u=0$ in
Eq.~\eqref{eq:G} gives
\begin{equation}
1-h^2\left(1+144\eps^2\right)=0,
\end{equation}
and the physical root is therefore
\begin{equation}
h(0)=\frac{1}{\sqrt{1+144\eps^2}},
\qquad
\left.
\partial_hG
\right|_{u=0,\,h=h(0)}
=
-2\sqrt{1+144\eps^2}\neq0,
\end{equation}
for every finite real $\eps$. The throat root consequently remains simple,
so the limiting branch point must occur away from $u=0$.

For real $\eps$, the double root equations can be reduced to an
algebraic equation for the critical ratio
$\gamma_\ast=|m/Q|$, which is explicitly written as 
\begin{equation}
348913664\,\gamma_\ast^6
-18903040\,\gamma_\ast^4
-75200\,\gamma_\ast^2
-125=0.
\label{eq:gammaPolynomial}
\end{equation}
The unique positive root is
\begin{equation}
\gamma_\ast\simeq0.24,
\end{equation}
and numerical continuation from $\eps=0$ shows that this is the first branch
point reached by the physical solution. The branch therefore remains
well defined throughout the wormhole for
\begin{equation}
\left|\frac{m}{Q}\right|<\gamma_\ast ,
\label{eq:numericalRange}
\end{equation}
and the corresponding branch point is located at
\begin{equation}
u_\ast\simeq0.99,
\qquad
h_\ast\simeq1.22+0.30\,i,
\label{eq:criticalPoint}
\end{equation}
which can be confirmed by numerically solving 
$G=\partial_hG=0$. 

%The critical value belongs to the classical theory defined by thetruncated action
Notice that the Gauss-Bonnet term introduces a minimal areal radius for the throat. Indeed, using $
 |Q|=2\sqrt{3}\,r_0^2$
together with $|m/Q|<\gamma_\ast$, we obtain
\begin{equation}
 r_0^2>
 \frac{|m|}{2\sqrt{3}\,\gamma_\ast}
 \simeq1.2\,|m|.
\end{equation}
Thus, at nonzero $m$,  asymptotically flat wormholes have a throat whose areal radius is bounded below, or in other words, such wormholes exist only if their throat radius is larger than $\sim1.1\sqrt{|m|}$. 
At the limiting point the expansion parameter $|m/Q|$ reaches $\gamma_\ast$, which should be understood as
the boundary of the truncated classical equations rather than as a
reliable ultraviolet cutoff. 
The Gauss-Bonnet correction is therefore no longer parametrically small
near the limiting point, and  thus
higher-order curvature operators cannot be assumed to remain suppressed
there.

\section{The KSW criterion}
\label{sec:ksw}

A complex solution is relevant only if its metric lies in the allowable domain of the gravitational path integral. For metric eigenvalues $\lambda_i$ the KSW condition~\cite{Kontsevich:2021dmb,Witten:2021nzp} is
\begin{equation}
 \sum_{i=1}^4|\Arg\lambda_i|<\pi,
 \label{eq:KSW}
\end{equation}
where the principal branch of the argument is understood. 

For the metric \eqref{eq:metricU} the three angular eigenvalues are positive and real. The only nontrivial phase belongs to the radial eigenvalue and obeys
\begin{equation}
 \Arg\lambda_u=-2\Arg h(u).
\end{equation}
The condition therefore reduces to
\begin{equation}
 2|\Arg h(u)|<\pi.
 \label{eq:KSW_h}
\end{equation}
Away from the throat this is the equivalent condition $|\Arg F|<\pi$. At the throat it must be evaluated in the regular coordinate $u$ since the zero of $F$ only reflects the failure of the areal radius as a coordinate there.

To show that the KSW condition is satisfied, let us suppose that the physical branch reaches
the KSW boundary away from the throat. Then $h$ should be purely imaginary
and, since $f=uh$ with real $u$, we may write
$
f=iy$, with real $y\neq 0$ so that \eqref{eq:alg} is written as
\begin{equation}
\begin{aligned}
0={}&
\left(12r^4-Q^2\right)\left(1+y^2\right)
\\
&+y^2\left[
(Q+16my)^2
+576m^2
+896m^2y^2
+320m^2y^4
\right]. 
\end{aligned}
\label{eq:noImaginaryF}
\end{equation}
Since for  $r\geq r_0$ one has $12r^4-Q^2\geq0$,  the first term in
\eqref{eq:noImaginaryF} is nonnegative, while the second is
strictly positive for $y\ne0$. The equation cannot be satisfied, so the
physical branch cannot cross the imaginary $h$ axis away from the throat.
At the throat $h(0)$ is explicitly real and positive. Since the branch
starts at $h=1$ and varies continuously, it follows that 
$\mathrm{Re}h(u)>0$
throughout its domain of existence. Consequently,
\begin{equation}
2|\Arg h(u)|<\pi,
\end{equation}
and the KSW condition is satisfied throughout the physical branch. The
solution therefore terminates at the algebraic branch point rather than at the KSW
boundary. 

\section{The scalar displacement and the imaginary distance bound}
\label{sec:scalar}

The scalar equation on the physical branch in the $u$-coordinates and the parametrization \eqref{eq:hdefinition} becomes
\begin{equation}
\frac{\dd\phi}{\dd u}
=
\frac{
iQ-8muh(u)\left[u^2h(u)^2-3\right]
}{
2r_0^2\sqrt{1-u^2}\,h(u)
}.
\label{eq:scalarU}
\end{equation}
This expression is finite at the throat and has only an integrable
$(1-u^2)^{-1/2}$ behavior at the two asymptotic ends. Since $h(u)$
remains finite and nonzero before the branch point, the scalar approaches
a finite value in each asymptotic region. In addition, the symmetry \eqref{eq:hsymmetry} implies
\begin{equation}
\frac{\dd\phi}{\dd u}(-u)
=
-\left(\frac{\dd\phi}{\dd u}(u)\right)^*,
\end{equation}
and choosing $\phi(0)=0$ we get that
\begin{equation}
\phi(-u)=\phi(u)^*.
\label{eq:phiSymmetry}
\end{equation}
Thus, the real part of the scalar is even and its imaginary part is odd, and therefore the
two endpoint values satisfy $\phi_-=\phi_+^*$.

For later use the expansion of the physical root for the $h$-function  is
\begin{equation}
\begin{aligned}
h(u)
={}&1+8i\eps u(1-u^2)
\\
&+\eps^2(1-u^2)
\left(
-72-16u^2+120u^4
\right)
+O(\eps^3),
\end{aligned}
\label{eq:hExpansion}
\end{equation}
which, when substitution into \eqref{eq:scalarU} gives
\begin{equation}
\begin{aligned}
\phi(u)
={}&i\frac{Q}{2r_0^2}\arcsin u
\\
&-\frac{4m}{r_0^2}
\left[
2\sqrt{1-u^2}
+(1-u^2)^{3/2}
-3
\right]
+O\!\left(\frac{m^2}{Q^2}\right).
\end{aligned}
\label{eq:phiExpansion}
\end{equation}
The zeroth order term reproduces the Einstein profile, while the leading
Gauss-Bonnet correction is real and even. It approaches the same value
$12m/r_0^2$ at both ends and therefore does not contribute to their
separation. Continuing the perturbative solution to the next order gives for the asymptotic values $\phi_\pm$ of the scalar at the endpoints $u=\pm 1$, respectively
\begin{equation}
\begin{aligned}
\phi_\pm
={}&
\frac{12m}{r_0^2}
+O\!\left(\frac{m^3}{Q^2r_0^2}\right)
\\
&\pm i\,\operatorname{sgn}(Q)\frac{\pi\sqrt3}{2}
\left[
1+90\frac{m^2}{Q^2}
+O\!\left(\frac{m^4}{Q^4}\right)
\right].
\end{aligned}
\label{eq:phiEndpoints}
\end{equation}
Since the endpoint values are complex conjugates, the natural continuation
of the quantity considered in \cite{Maldacena:2026jqd} is the half
endpoint displacement
\begin{equation}
d_{\wh}(Q,m)
=
\frac12|\phi_+-\phi_-|
=
|\mathrm{Im}\,\phi_+|,
\label{eq:dwhDefinition}
\end{equation}
which gives
\begin{equation}
d_{\wh}(Q,m)
=
\frac{\pi\sqrt3}{2}
\left[
1+90\frac{m^2}{Q^2}
+O\!\left(\frac{m^4}{Q^4}\right)
\right].
\label{eq:dwhExpansion}
\end{equation}
The Gauss-Bonnet interaction therefore increases the imaginary displacement
for a wormhole of finite size. This quantity measures the separation of the
endpoint values and should not be confused with the length of the full
complex scalar trajectory.

At fixed coupling the throat relation $Q^2=12r_0^4$ implies
\begin{equation}
\frac{m}{Q}\longrightarrow0
\qquad
\text{as}
\qquad
r_0\longrightarrow\infty,
\end{equation}
and therefore we have
\begin{equation}
\lim_{Q\to\infty}d_{\wh}(Q,m)
=
\frac{\pi\sqrt3}{2}
\label{eq:largeWormholeIDB}
\end{equation}
for every fixed and finite $m$. The finite size correction disappears in
the large wormhole limit relevant to the imaginary distance argument,  as also argued in Ref.
\cite{Maldacena:2026jqd}. Keeping $m/Q$ finite in this limit would
require sending the coupling $m$ to infinity together with $Q$ and would
therefore compare different theories.

The on-shell action provides a second important comparison with the
Einstein wormhole. We use the Dirichlet action with flat reference
subtraction at both asymptotic ends, so that $I_E$ measures the action
of the wormhole relative to the two disconnected flat geometries with
the same boundary data. 
When $m=0$ and $\Lambda=0$, the trace of Einstein's equations gives
\begin{equation}
 R-\frac12(\nabla\phi)^2=0,
\end{equation}
and hence the bulk action vanishes on shell. The complete two-sided
geometry has no boundary at the throat, while the flat reference
subtracted Gibbons--Hawking--York term
\cite{York:1972sj,Gibbons:1976ue} vanishes at both asymptotic ends, and 
therefore,
\begin{equation}
 I_E(Q,0)=0.
 \label{eq:einsteinActionZero}
\end{equation}
This result refers to the ensemble in which the scalar values are
fixed at the two ends. The fixed-charge action differs by the
corresponding scalar Legendre term.

For nonzero $m$, the complete Dirichlet action also contains the
scalar weighted Gauss-Bonnet boundary term
\cite{Myers:1987yn,Julie:2020vov}. As shown in
Appendix~\ref{app:action}, its contribution cancels against that of
the two flat reference geometries. After this subtraction, the field
equations reduce the on-shell action to
\begin{equation}
 I_E
 =
 -\frac{m}{16\pi G_N}
 \int\dd^4x\,\sqrt g\,\phi\,\GB.
 \label{eq:actionExact}
\end{equation}
The integral is convergent and real, and its leading contribution is
\begin{equation}
 I_E(Q,m)
 =
 \frac{3\sqrt3\,\pi}{2G_N}
 \left(64-15\pi\right)
 \frac{m^2}{|Q|}
 +
 O\left(\frac{m^4}{|Q|^3}\right).
 \label{eq:actionResult}
\end{equation}

The value \eqref{eq:actionResult} refers to the normalization $\phi(0)=0$,
i.e.\ to asymptotic data with $\mathrm{Re}\,\phi_\infty=12m/r_0^2$. Because of the topological shift of the action noted below
Eq.~\eqref{eq:action}, the subtracted action is not invariant under a constant
shift of the scalar, unlike in Einstein gravity. Since the wormhole has $\chi=0$ and each flat
reference geometry has $\chi=1$, we find
\begin{equation}
 I_E=I_{\rm wh}(Q)+\frac{4\pi m}{G_N}\,\mathrm{Re}\,\phi_\infty,
 \qquad
 I_{\rm wh}(Q)=-\frac{45\sqrt3\,\pi^2}{2G_N}\frac{m^2}{|Q|}
 +O\left(\frac{m^4}{|Q|^3}\right),
 \label{eq:actionSplit}
\end{equation}
where $\mathrm{Re}\,\phi_\infty=(\phi_++\phi_-)/2$ and $I_{\rm wh}$ is the
shift invariant action of the wormhole including its own boundary terms.
The wormhole contribution is negative at the order displayed,
while the topological term depends on the real scalar boundary
value. Their sum determines whether the wormhole is suppressed
relative to the disconnected geometries. At fixed $m$ and
$\mathrm{Re}\,\phi_\infty$, the wormhole contribution vanishes
as $|Q|\to\infty$, leaving a constant topological term. This
charge independent contribution changes the relative weighting
of topologies without changing the large charge convergence
threshold.

\section{A negative cosmological constant}
\label{sec:adsdistance}

It is interesting to ask whether the same construction extends to EAdS,
or hyperbolic space, with a negative cosmological constant
\begin{equation}
 \Lambda=-\frac{3}{\ell_{\rm AdS}^{2}}<0,
\end{equation}
and associated action
\begin{equation}
 I_{\rm bulk}[g,\phi,\Lambda]
 =-\frac{1}{16\pi G_N}
 \int \dd^4x\,\sqrt g
 \left[
 R-2\Lambda
 -\frac12\nabla_\mu\phi\nabla^\mu\phi
 +m\phi\GB
 \right].
 \label{eq:actionL}
\end{equation}
The scalar first integral is unchanged, while the radial constraint in the presence of a cosmological constant turns out to be
\begin{equation}
\begin{aligned}
0={}&12r^4-4\Lambda r^6-Q^2
-12\left(r^4+48m^2\right)f^2
\\
&+32imQf^3+1152m^2f^4-320m^2f^6.
\end{aligned}
\label{eq:adsConstraint}
\end{equation}
The areal radius $r_0$ of the throat is specified by the condition $f(r_0)=0$ which  lead to the relation
\begin{equation}
 Q^2
 =12r_0^4\left(1-\frac{\Lambda r_0^2}{3}\right)
 =12r_0^4(1+\eta),
 \qquad
 \eta\equiv\frac{r_0^2}{\ell_{\rm AdS}^2}.
 \label{eq:adsThroat}
\end{equation}
The Einstein connected branch again ends when two roots of the algebraic constraint coincide. The limiting value is now a function of the dimensionless throat size $\eta$ rather than a universal number as in the case of zero cosmological constant. 
The regular branch satisfies the bound
\begin{equation}
 \left|\frac{m}{Q}\right|<\gamma_\ast(\eta),
\end{equation}
where $\gamma_\ast(\eta)=1/160 s_\ast^{3/2}$ and $s_\ast$ satisfies 
the equation
\begin{align}
80s_\ast^3
\left(
2000s_\ast^3+800s_\ast^2+40s_\ast-11
\right)^2
=
9\eta^2(1+\eta)
\left(40s_\ast+7\right)^3. 
\label{eq:sast}
\end{align}
%For $\eta>0$, we select the positive solution of \eqref{eq:sast}
%that satisfies
%\begin{equation}
%2000s_\ast^3+800s_\ast^2+40s_\ast%-11>0.
%\end{equation}
%This condition follows from the unsquared branch point equations
%and selects the physical continuation of the flat space root.
We select the positive solution of \eqref{eq:sast} that reduces continuously to the flat
space result as $\eta\to0$, which gives for example the following values for $\eta=0$ and $\eta=1$
\begin{equation}
\gamma_\ast(0)\simeq0.24,
\qquad
\gamma_\ast(1)\simeq0.028.
\end{equation}
The negative cosmological constant therefore reduces the range of $|m/Q|$
over which the Einstein-connected branch exists. As in the asymptotically
flat case, this limiting value is a property of the truncated classical
equations and should not be interpreted as an ultraviolet cutoff.

The more important change concerns the scalar boundary behavior. Indeed, at large areal radius we find that $F=f^2$ is of  the form
\begin{equation}
 F(r)=k r^2+O(1),
 \qquad
 80m^2k^3+3k+\Lambda=0,
 \label{eq:F2}
\end{equation}
where the positive root 
$k$ gives an asymptotically hyperbolic metric.
The exact scalar first integral  is 
\begin{equation}
 \phi'(r)
 =
 -\frac{8m}{r^3}
 \left[f(r)^2-3\right]
 +\frac{iQ}{r^3f(r)},
 \label{eq:adsScalarExact}
\end{equation}
and using \eqref{eq:F2}, we find the asymptotic behaviour
\begin{equation}
 \phi'(r)
 =
 -\frac{8mk}{r}
 +O\left(\frac{m}{r^3}\right)
 +\frac{iQ}{\sqrt{k}\,r^4}
 +O\left(\frac{Q}{r^6}\right).
\label{eq:adsScalarDerivative}
\end{equation}
For $m\neq0$, the Gauss-Bonnet term dominates and gives
\begin{equation}
 \phi_\pm(r)
 =
 -8mk\log\left(\frac{r}{\ell_{\rm AdS}}\right)
 +\phi_\pm^{\rm fin}
 +O(r^{-2}).
 \label{eq:adsScalar}
\end{equation}
This logarithm has an important consequence. Any asymptotically AdS
geometry has $\GB\to24k^2\neq0$ near the boundary, so the scalar equation
\eqref{eq:sEOM} admits no constant solution and, for $\Lambda<0$, the coupling
$m\phi\GB$ acts as a tadpole. This is not tied to the spherical slicing,
since in Poincar\'e coordinates, $\dd s^2=\ell_{\rm AdS}^2(\dd z^2+\dd\vec x^{\,2})/z^2$,
one finds $\phi=(8m/\ell_{\rm AdS}^2)\log z+{\rm const}$ at leading order in
$m$. Hence, unlike in flat space where $\GB$ vanishes on the vacuum, $\phi$
is no longer a modulus and there is no family of EAdS vacua labelled by
$\phi_\infty$. In the dual description the radial dependence of $\phi$ translates into a
running of the coupling it sources, as in the holographic running couplings
of Refs.~\cite{Kehagias:1999tr,Gubser:1999pk}.  Here instead the logarithm is
forced by the bulk tadpole and corresponds to a constant beta function, of
magnitude $8|m|k$, closer to the logarithmic running of the cascading
solutions of Refs.~\cite{Klebanov:2000nc,Klebanov:2000hb}, so that the
boundary theory is not conformal for any value of this coupling.   The AdS
imaginary distance bound of Ref.~\cite{Maldacena:2026jqd}, formulated on a
space of conformal field theories, therefore does not apply directly.
However, since the beta function is real and, by the shift symmetry,
independent of $\phi$, only $\mathrm{Re}\,\phi$ runs while $\mathrm{Im}\,\phi$
is scale independent. The imaginary displacement defined below is thus
unambiguous, but it should be read as the imaginary part of a slowly running
coupling rather than as a distance on a conformal manifold.

When $m=0$ the logarithmic contribution vanishes, while the charge
contribution remains. Since $f_\pm(r)\sim\pm\sqrt{k}\,r$ in the two
exterior regions, one finds
\begin{equation}
 \phi'_{0,\pm}(r)
 =
 \pm\frac{iQ}{\sqrt{k}\,r^4}
 +O(r^{-6}),
\end{equation}
and therefore
\begin{equation}
 \phi_{0,\pm}(r)
 =
 \phi_\pm
 \mp\frac{iQ}{3\sqrt{k}\,r^3}
 +O(r^{-5}).
\end{equation}
The Einstein scalar is therefore nonconstant and purely imaginary, while it approaches a finite value at each asymptotic end. To measure the length
of a complex scalar profile, we use the positive Hermitian line element
on the complexified scalar plane
\begin{equation}
 \dd\ell_{\rm H}^2
 =
 \dd\phi\,\dd\bar\phi
 =
 |\dd \phi|^2,
\label{eq:hermitianLineElement}
\end{equation}
and the Hermitian length traced by the scalar along one exterior region is
then
\begin{equation}  d_m(\eta) = \lim_{R\to\infty} \int_{r_0}^{R} \sqrt{\phi'(r)\bar\phi'(r)}\,\dd r = \lim_{R\to\infty} \int_{r_0}^{R} |\phi'(r)|\,\dd r. \label{eq:hermitianLength} \end{equation}

At $m=0$ the radial constraint can be solved exactly giving the four dimensional asymptotically AdS wormhole solution 
Ref.~\cite{Maldacena:2026jqd}
\begin{equation}
 F_0(r)
 =
 1+\frac{r^2}{\ell_{\rm AdS}^2}
 -\frac{Q^2}{12r^4}
 =
 1+\frac{r^2}{\ell_{\rm AdS}^2}
 -(1+\eta)\frac{r_0^4}{r^4}.
 \label{eq:adsEinsteinMetric}
\end{equation}
 The last term is negligible near the
conformal boundary but is essential in the interior, where it produces
the zero $F_0(r_0)=0$ associated with the wormhole throat.
The exact scalar first integral then gives
\begin{equation}
 d_0(\eta)
 =
 |Q|
 \int_{r_0}^{\infty}
 \frac{\dd r}{r^3\sqrt{F_0(r)}},
\end{equation}
which, in terms of 
$y=r_0^2/r^2$ and 
$Q^2=12r_0^4(1+\eta)$ can be written as
\begin{equation}
 d_0(\eta)
 =
 \sqrt3
 \int_0^1
 \frac{\dd y}{
 \sqrt{
 1-y^2+
 \dfrac{\eta}{1+\eta}(y^{-1}-1)
 }}.
 \label{eq:d0int}
\end{equation}
The integrand in \eqref{eq:d0int} is a monotonically decreasing
function of $\eta$, and therefore,  $d_0(\eta)$ decreases monotonically
from $\pi\sqrt3/2$ as $\eta\to0$ to $\pi/\sqrt3$ as
$\eta\to\infty$. Thus every finite AdS wormhole satisfies
\begin{equation}
 \frac{\pi}{\sqrt3}
 <
 d_0(\eta)
 <
 \frac{\pi\sqrt3}{2}.
\end{equation}
This large throat value is the four dimensional AdS imaginary distance
proposed in Ref.~\cite{Maldacena:2026jqd}.

For $m\neq0$ the scalar trajectory is no longer confined to the
imaginary axis. Using its asymptotic behavior we get
\begin{equation}
 d_m(\eta)
 =
 \lim_{R\to\infty}
 \left[
 8|mk|\log\left(\frac{R}{r_0}\right)
 +O(1)
 \right],
\end{equation}
which is diverging, and therefore,
the scalar trajectory  has infinite Hermitian length. This
does not prevent the subtracted endpoint values from having a finite
difference, but that difference is a regulated displacement rather than
the length of the scalar trajectory.
The common real logarithm nevertheless cancels between the two ends. The
conjugation symmetry between the two exterior regions also applies to the
finite parts and gives
$ \phi_-^{\rm fin}
 =
 {\phi_+^{\rm fin}}^*.$
One may therefore define the finite regulated endpoint displacement
\begin{equation}
 d_m^{\rm reg}(\eta)
 =
 \frac12
 \left|
 \phi_+^{\rm fin}
 -
 \phi_-^{\rm fin}
 \right|
 =
 \left|
 \operatorname{Im}\phi_+^{\rm fin}
 \right|.
 \label{eq:adsRegulatedDisplacement}
\end{equation}
This quantity is finite but it is not the length of the scalar trajectory.

Since the logarithmic subtraction is real, it does not affect the
imaginary part of the scalar. Using $y=r_0^2/r^2$, the scalar first
integral gives
\begin{equation}
 d_m^{\rm reg}(\eta)
 =
 \left|
 \int_0^1
 \operatorname{Im}
 \left[
 -\frac{4m}{r_0^2}(f^2-3)
 +\frac{iQ}{2r_0^2f}
 \right]\dd y
 \right|.
 \label{eq:adsRegulatedIntegral}
\end{equation}
For $Q>0$ the orientation can be chosen so that the quantity inside
the absolute value is positive. Expanding the physical root of the
algebraic constraint in the large-throat limit then gives
\begin{equation}
 \lim_{\eta\to\infty}d_m^{\rm reg}(\eta)
 =
 \sqrt3
 \int_0^1
 \frac{\sqrt y\,\dd y}{\sqrt{1-y^3}}
 +
 \frac{8\sqrt3\,m^2}{\ell_{\rm AdS}^4}
 \int_0^1
 \sqrt y\sqrt{1-y^3}
 \left(5y^3-1\right)\dd y+
 O\left(\frac{m^4}{\ell_{\rm AdS}^8}\right),
\label{eq:adsReg}
\end{equation}
from where we find 
\begin{equation}
 \lim_{\eta\to\infty}d_m^{\rm reg}(\eta)
 =
 \frac{\pi}{\sqrt3}
 \left[
 1+
 \frac{m^2}{\ell_{\rm AdS}^4}
 +
 O\left(\frac{m^4}{\ell_{\rm AdS}^8}\right)
 \right].
 \label{eq:adsRegD}
\end{equation}
The Gauss-Bonnet interaction therefore shifts the large throat regulated
displacement upward at quadratic order. This result concerns the finite
difference between the subtracted endpoint values and not the Hermitian
length of the scalar trajectory, which remains infinite.
We stress that the modulus is lifted at first order in $m/\ell_{\rm AdS}^2$,
through the coefficient of the logarithm, while the shift in
\eqref{eq:adsRegD} is of second order. The result is therefore meaningful in
the regime $|m|\ll\ell_{\rm AdS}^2$, where $\phi$ is approximately marginal
over a range of scales $\Delta\log\mu\lesssim\ell_{\rm AdS}^2/(8|m|)$.

To relate the regulated displacement to the fixed charge action, we
consider the variation of the complete Dirichlet action $I_D$, where 
the subscript $D$ indicates that the induced metric and the scalar
value are fixed at the asymptotic boundaries. The Dirichlet action $I_D$ 
consists of the bulk action \eqref{eq:actionL}, the
Gibbons--Hawking--York term, the Gauss-Bonnet boundary term and the
local AdS counterterms.

For the spherical ansatz,
the on-shell variation with
respect to the scalar boundary value tuens out to be
\begin{equation}
 \delta I_D
 %\right|_{\rm on\ shell}
=
 \frac{\Omega_3}{16\pi G_N}
 \bigg[
 \Big{(}
 r^3f\phi'
 +8m\left(f^3-3f\right)
 \Big{)}
 \delta\phi
 \bigg]_-^+,
 \label{eq:DirichletBoundaryVariation}
\end{equation}
where $\Omega_3=2\pi^2$ is the volume of the three-sphere.
The expression multiplying $\delta\phi$ is independent of $r$ by the
scalar field equation. 
Indeed, the conserved  scalar canonical momentum is \begin{equation}
 p_\phi
 =
 \frac{\Omega_3}{16\pi G_N}
 \left[
 r^3f\phi'
 +8m\left(f^3-3f\right)
 \right]
 =i q, \qquad q=
 \frac{\Omega_3}{16\pi G_N}Q,
 \label{eq:adsScalarMomentum}
\end{equation}
where the scalar field equation has been used.
At a finite radial cutoff, the on-shell variation of the two-sided
Dirichlet action is then
\begin{equation}
 \delta I_D
 =
 p_\phi\delta\phi_+
 -p_\phi\delta\phi_-,
 \label{eq:DirichletVariationCutoff}
\end{equation}
where the relative minus sign follows from the opposite orientations of the
two asymptotic boundaries.

The coefficient of the logarithmic term in \eqref{eq:adsScalar} depends
only on $m$ and $\ell_{\rm AdS}$ and is independent of $Q$ so that it is constant along the family of solutions at fixed $m$ and
$\ell_{\rm AdS}$. After subtracting the common logarithmic term and
removing the cutoff, Eq.~\eqref{eq:DirichletVariationCutoff} becomes
\begin{equation}
 \delta I_D^{\rm ren}
 =
 p_\phi\,
 \delta\left(
 \phi_+^{\rm fin}-\phi_-^{\rm fin}
 \right).
 \label{eq:renormalizedDirichletVariation}
\end{equation}
Defining the full regulated imaginary displacement by
\begin{equation}
 \Delta\tau_{\rm fin}
 =
 -i\left(
 \phi_+^{\rm fin}-\phi_-^{\rm fin}
 \right)
 =
 2d_m^{\rm reg},
 \label{eq:regulatedFullDisplacement}
\end{equation}
where the orientation is chosen so that
$\Delta\tau_{\rm fin}>0$, and using $p_\phi=iq$, we obtain
\begin{equation}
 \delta I_D^{\rm ren}
 =
 -q\,\delta\Delta\tau_{\rm fin}.
 \label{eq:renormalizedDirichletVariation2}
\end{equation}
The fixed charge action is the Legendre transform~\cite{Maldacena:2026jqd}
\begin{equation}
 I_q^{\rm ren}
 =
 I_D^{\rm ren}
 +q\Delta\tau_{\rm fin},
 \label{eq:fixedChargeAction}
\end{equation}
so that its variation  satisfies
\begin{equation}
 \delta I_q^{\rm ren}
 =
 \Delta\tau_{\rm fin}\,\delta q,
\end{equation}
and hence
\begin{equation}
 \frac{\partial I_q^{\rm ren}}{\partial q}
 =
 \Delta\tau_{\rm fin}(q).
\end{equation}
Then, in the large-charge limit we get 
\begin{equation}
 \lim_{q\to+\infty}
 \frac{\partial I_q^{\rm ren}}{\partial q}
 =
 \frac{2\pi}{\sqrt3}
 \left[
 1+\frac{m^2}{\ell_{\rm AdS}^4}
 +O\!\left(\frac{m^4}{\ell_{\rm AdS}^8}\right)
 \right].
\label{eq:adsFixedChargeDer}
\end{equation}
This relation extends the Einstein fixed charge argument of
Refs.~\cite{Maldacena:2026jqd,DiUbaldo:2026rly,Rey:2026xtn} to the present
solutions with logarithmic scalar running. The limiting
displacement defines a regulated threshold, whose interpretation
assumes that the corresponding renormalized fixed charge sum
provides the appropriate semiclassical description.

\section{Conclusion}
\label{sec:conclusion}

We have examined how a linear scalar coupling to the four dimensional Gauss-Bonnet invariant modifies the complex wormholes underlying the imaginary distance proposal of Ref.~\cite{Maldacena:2026jqd}. The first effect of this interaction is qualitative as the scalar can no longer remain purely imaginary while the metric remains real. The Einstein wormhole nevertheless continues to a complex saddle whose two exterior regions are related by complex conjugation and joined through a smooth throat. This continuation exists only over a finite range of the Gauss-Bonnet coupling and terminates at an algebraic branch point. Throughout this range the metric satisfies the KSW allowability condition. The obstruction to extending the solution therefore comes from the field equations and not from the KSW boundary.

For asymptotically flat wormholes, the Gauss-Bonnet interaction changes both the scalar endpoint displacement and the on-shell action at finite throat size. The displacement increases relative to its Einstein value. The reference subtracted action splits into a negative, shift invariant wormhole contribution and a topological term proportional to $m\,\mathrm{Re}\,\phi_\infty$, so that whether finite wormholes are suppressed depends on the asymptotic data. The size dependent corrections disappear when the throat becomes large at fixed coupling. The main conclusion is therefore that the Gauss-Bonnet term modifies finite wormholes but does not shift the asymptotically flat imaginary distance threshold selected by arbitrarily large wormholes.

The situation in EAdS now is different. A nonzero Gauss-Bonnet coupling produces logarithmic scalar running near each conformal boundary. The positive Hermitian length of the scalar trajectory is consequently infinite, so the original imaginary distance bound cannot be extended directly as a bound on this length. However, the common logarithmic contribution cancels between the two ends and leaves a finite regulated endpoint displacement. Its large throat limit is shifted relative to the Einstein result and is related to the large charge behavior of the renormalized fixed charge action. Because the Gauss-Bonnet coupling lifts the scalar modulus for $\Lambda<0$, the dual coupling runs and the boundary theory is not conformal. Since only the real part runs, the displacement is scale independent and provides a natural AdS threshold, with the caveat that it refers to a running coupling rather than to a conformal manifold and relies on the convergence assumptions entering the fixed charge sum.

Several questions remain open. The fluctuation spectrum should be determined in order to establish the number and nature of negative modes. It is also necessary to identify whether these complex solutions lie on integration cycles connected to the original gravitational path integral. A complete holographic renormalization analysis would further
clarify the boundary interpretation of the running scalar and
the associated fixed charge ensemble. Finally, including higher curvature operators and more general scalar couplings would determine which of the present results survive beyond the four-derivative truncated effective action.

\medskip

\medskip

\noindent
\centerline{
{\bf Acknowledgments} } 
\vskip 0.2cm
\noindent
A.R.  acknowledges support from the Swiss National Science Foundation (project number CRSII5\_213497). Large language models (Claude and ChatGPT) were used for editorial assistance and language polishing.

\vspace{8mm}

\noindent
{\bf \Large Appendix}

\appendix

\section{The on-shell action}
\label{app:action}

The trace of the Gauss-Bonnet contribution in \eqref{eq:Hmn} is
\begin{equation}
 g^{\mu\nu}\mathcal H_{\mu\nu}
 =
 2R\Box\phi
 -4R^{\mu\nu}\nabla_\mu\nabla_\nu\phi
 =
 -4G^{\mu\nu}\nabla_\mu\nabla_\nu\phi.
\end{equation}
Taking the trace of the metric equation and using
$\Box\phi=-m\GB$ gives
\begin{equation}
 R-\frac12(\nabla\phi)^2
 =
 -4mR^{\mu\nu}\nabla_\mu\nabla_\nu\phi
 -2m^2R\GB.
 \label{eq:traceRelation}
\end{equation}
The contracted Bianchi identity gives
\begin{equation}
 \int\dd^4x\,\sqrt g\,
 R^{\mu\nu}\nabla_\mu\nabla_\nu\phi
 =
 -\frac m2
 \int\dd^4x\,\sqrt g\,R\GB
\end{equation}
up to boundary terms. Those terms vanish because the fields approach their asymptotic values sufficiently rapidly at both ends. The two terms proportional to $m^2R\GB$ then cancel. Including the overall normalization in \eqref{eq:action} leaves
\begin{equation}
 I_{\rm bulk}
 =
 -\frac{m}{16\pi G_N}
 \int\dd^4x\,\sqrt g\,\phi\GB.
 \label{eq:actionExact-append}
\end{equation}
which proves \eqref{eq:actionExact}.

Using  the standard background subtraction prescription for
asymptotically flat gravity
\cite{Hawking:1995fd},
the gravitational Dirichlet action contains  the flat reference
subtracted Gibbons--Hawking--York term~\cite{York:1972sj,Gibbons:1976ue}. The latter   at a cutoff surface $r=R$ turns out to be
\begin{equation}
K=\frac{3\sqrt F}{R},
\qquad
K_0=\frac{3}{R},
\qquad
\sqrt\gamma\left(K-K_0\right)
=
3R^2\left(\sqrt F-1\right)
=
O(R^{-2}),
\end{equation}
so that its contribution vanishes at both asymptotic ends. On the other hand, the scalar Gauss-Bonnet interaction 
requires also the additional Dirichlet
boundary term
\cite{Myers:1987yn,Julie:2020vov}
\begin{equation}
I_{\partial}
=
-\frac{m}{4\pi G_N}
\int_{\partial M}
\dd^3x\,\sqrt\gamma\,\phi
\left(
J-2\widehat G_{ij}K^{ij}
\right).
\label{eq:GBBoundaryTerm}
\end{equation}
Let us now define 
$f_\perp=n^\mu\partial_\mu r$
using the outward normal $n^\mu$, which approaches one at both
asymptotic ends. On a spherical cutoff surface of radius $R$ we find 
\begin{equation}
K_i{}^j
=
\frac{f_\perp}{R}\delta_i{}^j,
\qquad
J
=
-\frac{2f_\perp^3}{R^3},
\qquad
\widehat G_{ij}K^{ij}
=
-\frac{3f_\perp}{R^3},
\end{equation}
and therefore
\begin{equation}
\int_{S_R^3}
\dd^3x\,\sqrt\gamma
\left(
J-2\widehat G_{ij}K^{ij}
\right)
=
-2\Omega_3
\left(
f_\perp^3-3f_\perp
\right)
\longrightarrow
4\Omega_3.
\end{equation}
 Using
\begin{equation}
\phi_++\phi_-
=
\frac{24m}{r_0^2}
+
O\left(\frac{m^3}{Q^2r_0^2}\right)
\end{equation}
and $r_0^2=|Q|/(2\sqrt3)$ we find
\begin{equation}
I_{\partial}
=
-\frac{96\sqrt3\,\pi}{G_N}
\frac{m^2}{|Q|}
+
O\left(\frac{m^4}{|Q|^3}\right).
\label{eq:GBBoundaryValue}
\end{equation}
We apply the same flat reference subtraction to this boundary term.
The flat reference geometries have the same induced metric and scalar
values as the two asymptotic ends of the wormhole. Their boundary
contribution is therefore equal to \eqref{eq:GBBoundaryValue} and
cancels from the subtracted action. Since the subtracted
Gibbons--Hawking--York term also vanishes, the complete action is
given by the bulk contribution,
\begin{equation}
 I_E=I_{\rm bulk}.
\end{equation}
For the perturbative solution we have
\begin{equation}
 m\int\dd^4x\,\sqrt g\,\phi\GB
 =
 24\sqrt3\,\pi^2(15\pi-64)
 \frac{m^2}{|Q|}
 +
 O\left(\frac{m^4}{|Q|^3}\right),
\end{equation}
and therefore we find that $I_E$ is given by 
\begin{equation}
 I_E
 =
 \frac{3\sqrt3\,\pi}{2G_N}
 \left(64-15\pi\right)
 \frac{m^2}{|Q|}
 +
 O\left(\frac{m^4}{|Q|^3}\right).
\end{equation}
Here $I_E$ is the action relative to the two disconnected flat
reference geometries and is real and positive at this order
for $\phi(0)=0$. Its relation to the shift invariant wormhole
action is given in Eq.~\eqref{eq:actionSplit}.

\bibliographystyle{JHEP}
\bibliography{biblio}

\end{document}